\documentclass[12pt]{article}
\usepackage{times}
\usepackage{fullpage}
\usepackage{amsmath}
\usepackage{amsbsy}
\usepackage{amsfonts}
\usepackage{supertabular}
\usepackage{graphicx}
\usepackage{subfigure}
\usepackage{natbib}
\usepackage{epsfig}
\usepackage{xspace}

\def\asinh{{\rm asinh}}
\def\bcc{bcc\xspace}
\def\fcc{fcc\xspace}

\def\d0{{\it d}}

 \def\bb{{\bf b}}

\newcommand{\ccc}{{\c C}a{\u g}{\i}n}

\title{\bfseries{\Large 
A Multiscale Approach for Modeling Crystalline Solids}}

\author{
\normalsize{\bf A. M. Cuiti\~no $^1$, L. Stainier $^2$, Guofeng Wang $^3$, Alejandro Strachan $^3$,}\\ 
\normalsize{\bf Tahir \ccc $^3$, William A. Goddard, III $^3$, and M. Ortiz $^4$} \\
\small{$^1$ Department of Mechanical and Aerospace Engineering} \\
\small{Rutgers University,  Piscataway, NJ 08854, USA.} \medskip \\
\small{$^2$ Laboratoire de Techniques A\'eronautiques et Spatiales} \\
\small{        University of Li\`ege, 4000 Li\`ege, Belgium.} \medskip \\
\small{$^3$ Materials and Process Simulation Center,Beckman Institute (139-74)}\\
\small{ California Institute of Technology, Pasadena, CA 91125, USA.}
\medskip \\
\small{$^4$ Graduate Aeronautical Laboratories}\\
\small{        California Institute of Technology, Pasadena, CA 91125, USA.}}

\begin{document}
\maketitle
\begin{abstract}
In this paper we present a modeling approach to bridge the atomistic
with macroscopic scales in crystalline materials. The methodology
combines identification and modeling of the controlling unit processes
at microscopic level with the direct atomistic determination of
fundamental material properties. These properties are computed using a
many body Force Field derived from ab initio quantum-mechanical
calculations. This approach is exercised to describe the mechanical
response of high-purity Tantalum single crystals, including the effect
of temperature and strain-rate on the hardening rate. The resulting
atomistically informed model is found to capture salient features of
the behavior of these crystals such as: the dependence of the initial
yield point on temperature and strain rate; the presence of a marked
stage I of easy glide, specially at low temperatures and high strain
rates; the sharp onset of stage II hardening and its tendency to shift
towards lower strains, and eventually disappear, as the temperature
increases or the strain rate decreases; the parabolic stage II
hardening at low strain rates or high temperatures; the stage II
softening at high strain rates or low temperatures; the trend towards
saturation at high strains; the temperature and strain-rate dependence
of the saturation stress; and the orientation dependence of the 
hardening rate.
\end{abstract}
\newpage
\section{Introduction}
\label{introduction}
This paper is concerned with the development of a multiscale modeling
approach for advanced materials such as high-purity \bcc single
crystals. The present approach is aligned with the current {\it divide
and conquer} paradigm in micromechanics (see, e.~g., \cite{BulatovKubin1998,
Phillips1998, CampbellFoilesHuang1998, PhillipsRodneyShenoy1999,
MoriartyXuSoderlind1999, Baskes1999}. This paradigm first identifies 
and models the controlling unit process at microscopic scale. Then, 
the energetics and dynamics of these mechanisms are quantified by means
of atomistic modeling. Finally, the macroscopic driving force is
correlated to macroscopic response via microscopic modeling. This last
step involves two stages, {\it localization} of the macroscopic
driving force into unit-process driving forces and {\it averaging} of
the contribution of each unit process into the macroscopic response.

We show in this article that the meticulous application of this
paradigm renders truly predictive models of the mechanical behavior of
complex systems. In particular we predict the hardening of Ta single
crystal and its dependency for a wide range of temperatures and strain
rates. The feat of this approach is that predictions from these
atomistically informed models recover most of the macroscopic
characteristic features of the available experimental data, without a
priori knowledge of such experimental tests. This approach then
provides a procedure to forecast the mechanical behavior of material
in extreme conditions where experimental data is simply not available
or very difficult to collect. 

A crucial step in this approach is the appropriate selection and
modeling of the unit processes. These models supply the link between
the atomic and meso scale by identifying and correlating the relevant
material properties, susceptible of atomistic determination such as
energy formation for defects, with the corresponding driving
forces. In this case, we specifically consider the following unit
processes: double-kink formation and thermally activated motion of
kinks; the close-range interactions between primary and forest
dislocation, leading to the formation of jogs; the percolation motion
of dislocations through a random array of forest dislocations
introducing short-range obstacles of different strengths; dislocation
multiplication due to breeding by double cross-slip; and dislocation 
pair-annihilation.  

A set of material parameters is then obtained from the modeling and
identification stage, which is required to quantify the contribution
of each of the unit processes. 
We compute these materials properties using a combination of ab-initio
quantum mechanics (QM) and Force Field (FF) calculations. QM describes
the atomic interactions from first principles, i.e. with no input from
experiments; unfortunatelly QM methods
are computationally intensive and restricted to small systems, making 
QM calculations impractical to study most of the materials properties governing 
plasticity. Force Fields give the total energy of a system as a potential
energy function of the atomic positions and with Molecular Dynamics (MD)
allows the simulation of systems containing millions of atoms.
We used ab-initio quantum mechanical calculations (equations of state of
various crystalline phases, elastic constants, energetics of defects, etc.)
to develop a many body Force Field (FF) (named qEAM FF) for Tantalum. 
We use the qEAM FF with MD to calculate the
core energy of the $1/2a<111>$ screw dislocation, that of the edge
dislocation with Burgers vector $b=1/2 a <111>$ in (110) planes. We have
also calculated  the
formation energies and nucleation lengths of the kinks in $b=1/2 a <111>$
screw dislocations.

The organization of the paper follows the sequential stages of the
proposed approach. First, we provide a brief description of each of
the unit processes including the governing final equations. We then
identify and compute by atomistic means the corresponding material
properties. Finally, we compare the predictions against experimental
data. 

\section{Unit Processes}

Plastic deformation in metallic systems is the macroscopic
manifestation of dislocation activity. The resistance to the
dislocation motion, therefore, engenders the hardening properties
observed in this type of materials. It is then the complex interplay of
microscopic mechanisms controlling {\bf dislocation mobility, dislocation
interaction and dislocation evolution} which confers the macroscopic
constitutive properties. In the present approach, these controlling  
processes are considered to be {\it orthogonal} in the sense that are
weakly coupled with each other. The interaction among them is only 
established through the  uniqueness of the macroscopic driving force 
which are shared, via the localization process, by all the unit 
processes.

In this section, we introduce the set of controlling unit processes 
which have been identified for describing the mechanical response of
high-purity BCC single crystals, in particular for Tantalum. 
We also provide the final expression resulting from the the modeling
of each of these processes. A detailed description of the model,
including comparison with experimental data is given in
\cite{StainierCuitinoOrtiz2001}.

\begin{figure}
\centerline{ \hbox{
\subfigure[]
{\epsfig{file=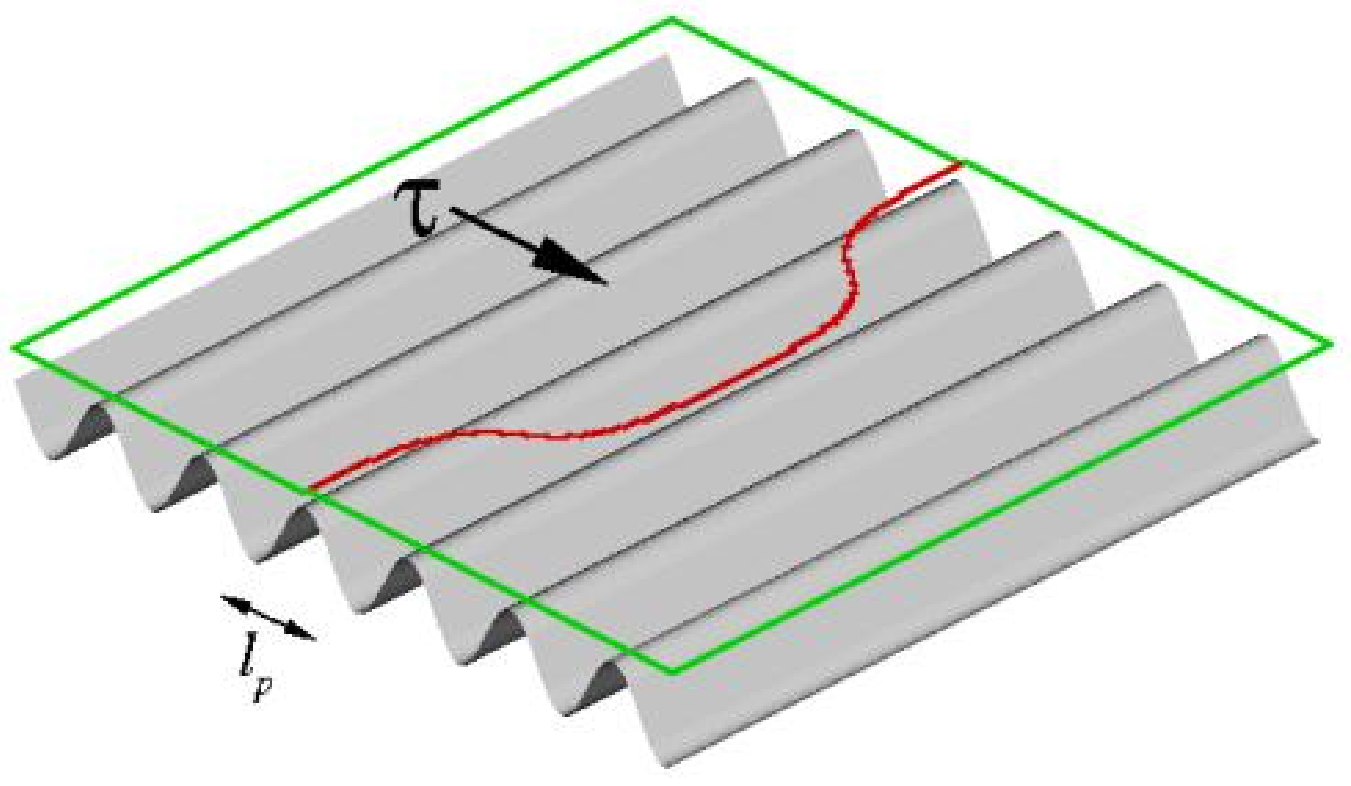,height=0.30\textheight}} 
\subfigure[]
{\epsfig{file=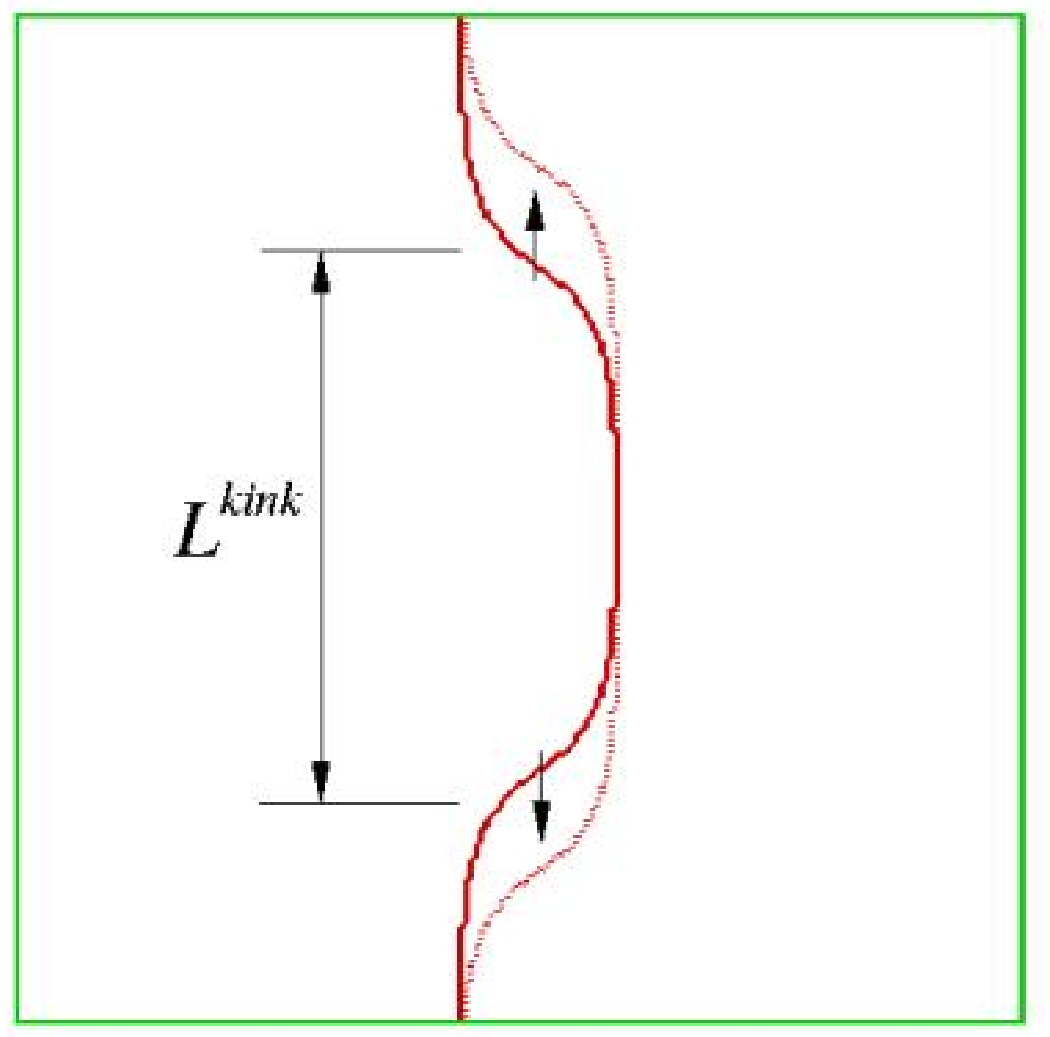,height=0.25\textheight}}}}
\caption{Schematic of the double-kink mechanism.}
\label{fig:kinkpair}
\end{figure}    

\subsection{Dislocation Mobility: Double-Kink Formation and Thermally
  Activated Motion of Kinks}

We consider the thermally activated motion of dislocations within an
\emph{obstacle-free} slip plane. Under these conditions, the motion
of the  dislocations is driven by an applied resolved shear stress
$\tau$ and is hindered by the
lattice resistance, which is weak enough that it may be overcome
by thermal activation. The lattice resistance is presumed to be
well-described by a Peierls energy function, which assigns an
energy per unit length to dislocation segments as a function of
their position on the slip plane.

In \bcc crystals, the core of screw dislocation segments relaxes
into low-energy non-planar configurations
\cite{DuesberyVitekBowen1973, Vitek1976, Vitek1992,
XuMoriarty1996, DuesberyVitek1998, MoriartyXuSoderlind1999,
Ismail-beigiArias2000, WangStrachanCaginGoddard2000}.  This
introduces deep valleys into the Peierls energy function aligned
with the Burgers vector directions and possessing the periodicity
of the lattice.  At low temperatures, the dislocations tend to
adopt low-energy configurations and, consequently, the dislocation
population predominantly consists of long screw segments. In order
to move a screw segment normal to itself, the dislocation core
must first be constricted, which requires a substantial supply of
energy. Thus, the energy barrier for the motion of screw segments,
and the attendant Peierls stress, may be expected to be large, and
the energy barrier for the motion of edge segments to be
comparatively smaller. For instance, Duesbery and Xu
\cite{DuesberyXu1998} have calculated the Peierls stress for a
rigid screw dislocation in Mo to be 0.022$\mu$, where $\mu$ is the
$\langle 111 \rangle$ shear modulus, whereas the corresponding
Peierls stress for a rigid edge dislocation is 0.006$\mu$, or
about one fourth of the screw value. This suggests that the
rate-limiting mechanism for dislocation motion is the thermally
activated motion of kinks along screw segments (\cite{Hirsch1960,
SeegerSchiller1962, hirth:1968}).
 
\begin{figure}
\centerline{\epsfig{file=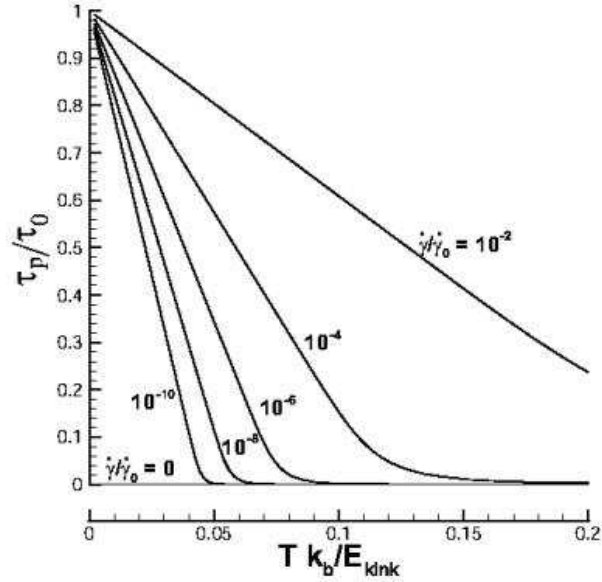,width=0.9\textwidth}}
\vglue -0.1\textwidth
\caption{Temperature dependence of the effective Peierls stress
for various strain rates. Note that the typical order of magnitude
of $\dot{\gamma}^{\rm kink}_0 = 10^6\:s^{-1}$.}
\label{fig:thermal}
\end{figure}

At sufficiently high temperatures and under the application of a
resolved shear stress $\tau > 0$, a double-kink may be nucleated with
the assistance of thermal activation (e.~g.,
\cite{HirthHoagland1993, XuMoriarty1998, MoriartyXuSoderlind1999},
and the subsequent motion of the kinks causes the screw segment to
effectively move forward, Fig.~\ref{fig:kinkpair}. Under this
conditions the following expression for the effective temperature and
strain-rate dependent Peierls $\tau_p$ is obtained:
\begin{equation}
\tau_P = \frac{\tau_0}{\beta E^{\rm
kink}} \asinh\left(\frac{\dot{\gamma}}{\dot{\gamma}^{\rm kink}_0}{\rm
e}^{\beta E^{\rm kink}}\right)
\label{YieldStressDependence}
\end{equation}
where the effective Peierls stress is given by
\begin{equation}\label{tauzero}
\tau_0 = \frac{E^{\rm kink}}{b L^{\rm kink} l_P}
\end{equation}
and the reference strain is defined as
\begin{equation}
\dot{\gamma}^{\rm kink}_0 = 2 b \rho l_P \nu_D
\end{equation}
In the preceeding equations, $b$ is the
Burgers vector, $\rho$ is the dislocation density, $\beta = 1/k_B
T$, $k_B$ is Boltzmann's constant, $T$ is the absolute temperature,
and $\nu_D$ is the attempt frequency which may be identified with the
Debye frequency to a first approximation. Also, $l_P$ is the distance
between two consecutive Peierls valleys. For \bcc crystals, $l_P = \sqrt{2/3} a$ if the slip plane is
$\{110\}$, $l_P = \sqrt{2} a$, if the slip plane is $\{112\}$, and
$l_P = \sqrt{8/3} a$ if the slip plane is $\{123\}$, where $a$ is
the cubic lattice size \cite{SeegerHollang2000}. Finally, $E^{\rm
  kink}$ is the energy of formation
of a kink-pair and $L^{\rm kink}$ is the length of an incipient double
kink. The formation energy $E^{\rm kink}$ and the length $L^{\rm
  kink}$, which cannot be reliably estimated from elasticity since the
energy is composed mostly of core region, can, however, be accurately
computed by recourse to atomistic models as shown in
section~\ref{atomistic}. Modeling of this first unit process renders
the first 2 material properties amenable of atomistic calculations.

\begin{figure}
\vglue -0.1\textwidth
\centerline{ \epsfig{file=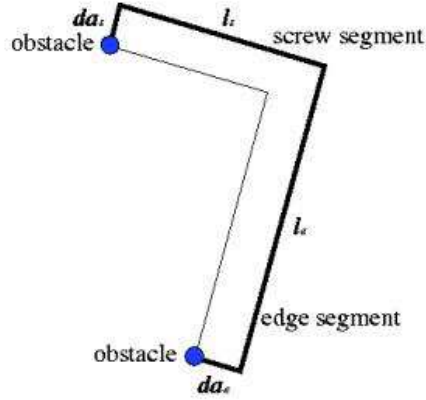,width=0.9\textwidth} }
\vglue -0.1\textwidth
\caption{Bow-out mechanism for a dislocation segment bypassing an
obstacle pair} \label{fig:segments}
\end{figure}

In Figure \ref{fig:thermal} the dependence of the effective Peierls  
stress on temperature and rate of deformation is illustrated.
The Peierls stress decreases ostensibly linearly up to a critical
temperature $T_c$, beyond which it tends to zero. These
trends are in agreement with the experimental observations of
Wasserb\"ach \cite{Wasserbach1986} and Lachenmann and Schultz
\cite{LachenmannShultz1970}. The critical temperature $T_c$
increases with the strain rate. In particular, in this model the
effect of increasing (decreasing) the strain rate has an analogous
effect to decreasing (increasing) the temperature, and vice-versa,
as noted by Tang {\it et al.} \cite{TangDevincreKubin1999}. In the
regime of very high strain-rates ($\dot{\gamma} > 10^5 \:
s^{-1}$), effects such as electron and phonon drag become
important and control the velocity of dislocations
\cite{suzuki:1991, brailsford:1969}.

\subsection{Dislocation Interactions: Obctacle-Pair Strength and   
Obstacle Strength}

In the forest-dislocation theory of hardening, the motion of
dislocations, which are the agents of plastic deformation in
crystals, is impeded by secondary --or `forest'-- dislocations
crossing the slip plane. As the moving and forest dislocations
intersect, they form jogs or junctions of varying strengths
\cite{baskes:1998, RodneyPhillips1999, PhillipsRodneyShenoy1999,
ShenoyKuktaPhillips2000, DannaBenoit1993, RheeZbibHirth1998,
HuangGhoniemDelaRubia1999, KubinDevincreTang1998,
ZbibDeLaRubiaRhee2000} which, provided the junction is
sufficiently short, may be idealized as point obstacles. Moving
dislocations are pinned down by the forest dislocations and
require a certain elevation of the applied resolved shear stress
in order to bow out and bypass the pinning obstacles.  
For the case of infinitely strong obstacles, the resistance of the
forest is provided by the strength of the obstacle pairs. This
obstacle pair strength is subsequently reduced by considering that 
point obstacles composing the pair can only provide a finite strength.
The processes imparting the pair-obstacle strength and obstacle
strength are described next

\begin{figure}
\vglue -0.1\textwidth
\centerline{
  \epsfig{file=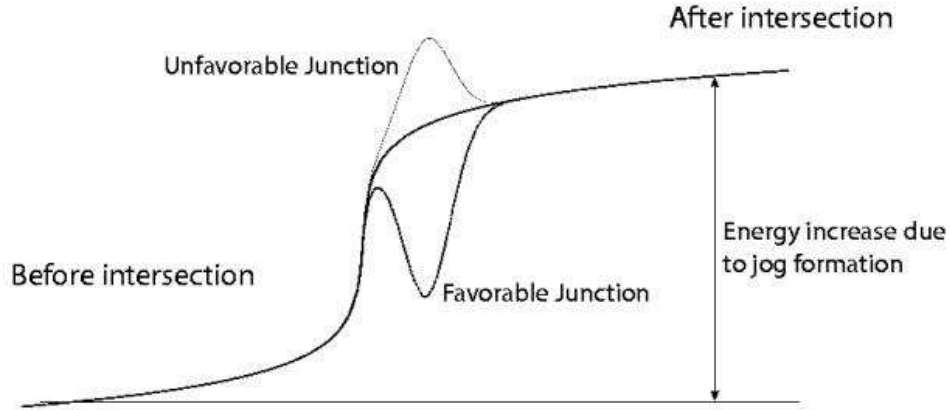,width=0.9\textwidth}
} 
\vglue -0.1\textwidth
\caption{Schematic of energy variation as a function of a
reaction coordinate during dislocation intersection and crossing.}
\label{fig:junction}
\end{figure}

\subsubsection{Obstacle-Pair Strength} 

We begin by treating the case of infinitely strong obstacles.  In this
case, pairs of obstacles pin down dislocation segments, which require a
certain threshold resolved shear stress $s$ in order to overcome
the obstacle pair. The lowest-energy configuration of unstressed 
dislocation segments spanning an obstacle pair is a step of the form shown 
as the thin line in
Fig.~\ref{fig:segments}.  Under these conditions, the bow-out
mechanism by which a dislocation segment bypasses an obstacle pair
may be expected to result in the configuration shown in
Fig.~\ref{fig:segments} (bold line).  If the edge-segment length
is $l_{e}$, a displacement $da_{e}$ of the dislocation requires a
supply of energy equal to $2 U^{\rm screw} da_{e} + b \tau_P^{\rm
edge} l_{e} da_{e}$ in order to overcome the Peierls resistance
$\tau_P^{\rm edge}$ and to extend the screw segments.  The
corresponding energy release is $b \tau l_{e} da_{e}$.  Similar
contributions result from a displacement $da_{s}$ of the
screw-segment of length $l_{s}$. Retaining dominant terms the
obstacle-pair strength is
\begin{equation}\label{TauC}
s = \tau_P^{\rm screw} + \frac{2 U^{\rm edge}}{b l_{s}}
\end{equation}
The obstacle-pair strength can be therefore estimated by quantifying 
$\tau_P$, $l_{s}$ and $U^{\rm edge}$. An expression for the Peierls
stress $\tau_P$ is given in Eq.~(\ref{YieldStressDependence}), the
distance between obstacles along the screw direction $l_{s}$ is
estimated by statistics assuming a random 
obstacle distribution and the core energy per unit length in the edge
direction $U^{\rm edge}$ is obtained by atomistic calculations
presented in the following sections. 

\subsubsection{Obstacle Strength}

In this section we proceed to estimate the obstacle strengths
which reduces the obstacle-pair strength described in the previuos 
section. The interaction between primary and secondary dislocations
may result in a variety of reaction products, including jogs and junctions
\cite{DannaBenoit1993, RheeZbibHirth1998, baskes:1998,
HuangGhoniemDelaRubia1999, TangDevincreKubin1999,
RodneyPhillips1999, PhillipsRodneyShenoy1999,
ZbibDeLaRubiaRhee2000, ShenoyKuktaPhillips2000}. Experimental
estimates of junction strengths have been given by Franciosi and
Zaoui \cite{franciosi:1982} for the twelve slip systems belonging
to the family of $\{111\}$ planes and $[110]$ directions in \fcc
crystals, and by Franciosi \cite{franciosi:1983} for the
twenty-four systems of types $\{211\}$ $[111]$ and $\{110\}$
$[111]$ in \bcc crystals.  The strength of some of these
interactions has recently been computed using atomistic and
continuum models \cite{baskes:1998, RodneyPhillips1999,
PhillipsRodneyShenoy1999, ShenoyKuktaPhillips2000}. Tang {\it et
al.} have numerically estimated the average strength of
dislocation junctions for Nb and Ta crystals
\cite{TangDevincreKubin1999}.

\begin{figure}
\vglue -0.1\textwidth
\centerline{
  \epsfig{file=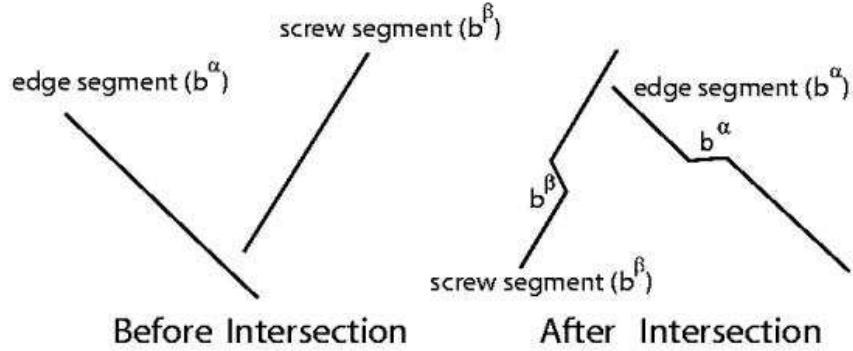,width=0.9\textwidth}
} 
\vglue -0.2\textwidth
\caption{Schematic of jog formation during dislocation
intersection} \label{fig:jogs}
\end{figure}

For purposes of the present theory, we specifically concern
ourselves with short-range interactions between dislocations which
can be idealized as point defects. For simplicity, we consider the
case in which each intersecting dislocation acquires a jog. The
energy of a pair of crossing dislocations is schematically shown
in Fig.~\ref{fig:junction} as a function of some convenient
reaction coordinate, such as the distance between the
dislocations. The interaction may be repulsive, resulting in an
energy barrier, or attractive, resulting in a binding energy,
Fig.~\ref{fig:junction}. In the spirit of an equilibrium theory,
here we consider only the final reaction product, corresponding to
a pair of jogged dislocations at infinite distance from each
other, and neglect the intermediate states along the reaction
path. In addition, we deduce the strength of the obstacles
directly from the energy supply required to attain the final
state, i.~e. the jog-formation energy. Despite the sweeping nature
of these assumptions, the predicted saturation strengths in
multiple slip are in good agreement with experiment (cf
Section~\ref{ComparisonExperiment}), which lends some empirical
support to the theory.

\begin{figure}
\vglue -0.2\textwidth
\centerline{
  \epsfig{file=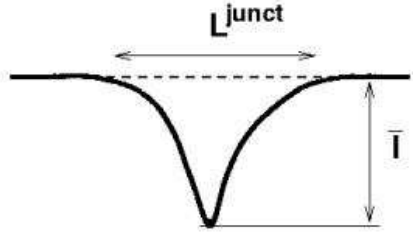,width=0.9\textwidth}
} 
\vglue -0.2\textwidth
\caption{Schematic of a dislocation line overcoming a junction}
\label{fig:junction2}
\end{figure}

We estimate the jog formation energy as follows. Based on energy
and mobility considerations already discussed, we may expect the
preponderance of forest dislocations to be of screw character, and
the mobile dislocation segments to be predominantly of edge
character. We therefore restrict our analysis to intersections
between screw and edge segments. The geometry of the crossing
process is schematically shown in Fig.~\ref{fig:jogs}. Each
dislocation acquires a jog equal to the Burgers vector of the
remaining dislocation. The energy expended in the formation of the
jogs may be estimated as

\begin{equation}
E^{jogs}_{\alpha\beta} \sim
\begin{cases}
b U^{\rm screw} \left[ 1 - r \cos\theta^{\alpha\beta} \right] & 
\text{if } \bb^\alpha = \bb^\beta \\
b U^{\rm screw} \left[ 2 r - \cos(\theta^{\beta\alpha}) - r
\cos\theta^{\alpha\beta} \right]   & \text{otherwise}
\end{cases}
\label{energies}
\end{equation}
where $r = U^{\rm edge}/U^{\rm screw}$ is the ratio of screw to
edge dislocation line energies. This ratio is computed by atomistic 
calculations presented in the next section, renders a value of $r =
1.77$ for Ta. The resulting jog formation energies for the complete 
collection of pairs of $\{211\}$ and $\{110\}$ dislocations are
tabulated in Table~\ref{tab:matprop3}.

\bigskip

\begin{table}
\centering
{\footnotesize{
\begin{tabular}
{ | l | l@{} l@{} l@{} l@{} l@{} l@{} l@{} l@{} l@{} l@{} l@{} l@{}
  l@{} l@{} l@{} l@{} l@{} l@{} l@{} l@{} l@{} l@{} l@{} l@{} |}
\hline
\space & {\bfseries A2\space} & {\bfseries A2'} & {\bfseries
A3\space} & {\bfseries A3'} &
     {\bfseries A6\space} & {\bfseries A6'} & {\bfseries B2\space} &
{\bfseries B2''} &
     {\bfseries B4\space} & {\bfseries B4'} & {\bfseries B5\space} &
{\bfseries B5'} &
     {\bfseries C1\space} & {\bfseries C1'} & {\bfseries C3\space} &
{\bfseries C3''} &
     {\bfseries C5\space} & {\bfseries C5''} & {\bfseries D1\space} &
{\bfseries D1''} &
     {\bfseries D4\space} & {\bfseries D4''} & {\bfseries D6\space} &
  {\bfseries D6''} \\ \hline
{\bfseries A2}   &  ---- &  1.0 &  1.0 &  1.0 &  1.0 &  1.0 &  1.5 &
1.5 &  1.5 &  1.5 &  1.5 &  1.5 &  2.4 &  2.4 &  2.4 &  2.4 &  2.4 &
2.4 &  2.4 &  2.4 &  2.4 &  2.4 &  2.4 &  2.4 \\
{\bfseries A2'}  &  1.0 &  ---- &  1.0 &  1.0 &  1.0 &  1.0 &  3.2 &
3.2 &  3.2 &  3.2 &  3.2 &  3.2 &  1.8 &  1.8 &  1.8 &  1.8 &  1.8 &
1.8 &  1.8 &  1.8 &  1.8 &  1.8 &  1.8 &  1.8 \\
{\bfseries A3} &  1.0 &  1.0 &  ---- &  1.0 &  1.0 &  1.0 &  2.4 &
2.4 &  2.4 &  2.4 &  2.4 &  2.4 &  1.5 &  1.5 &  1.5 &  1.5 &  1.5 &
1.5 &  2.4 &  2.4 &  2.4 &  2.4 &  2.4 &  2.4 \\
{\bfseries A3'}  &  1.0 &  1.0 &  1.0 &  ---- &  1.0 &  1.0 &  1.8 &
1.8 &  1.8 &  1.8 &  1.8 &  1.8 &  3.2 &  3.2 &  3.2 &  3.2 &  3.2 &
3.2 &  1.8 &  1.8 &  1.8 &  1.8 &  1.8 &  1.8 \\
{\bfseries A6}  &  1.0 &  1.0 &  1.0 &  1.0 &  ---- &  1.0 &  2.4 &
2.4 &  2.4 &  2.4 &  2.4 &  2.4 &  2.4 &  2.4 &  2.4 &  2.4 &  2.4 &
2.4 &  1.5 &  1.5 &  1.5 &  1.5 &  1.5 &  1.5 \\
{\bfseries A6'} &  1.0 &  1.0 &  1.0 &  1.0 &  1.0 & ---- &  1.8 &
1.8 &  1.8 &  1.8 &  1.8 &  1.8 &  1.8 &  1.8 &  1.8 &  1.8 &  1.8 &
1.8 &  3.2 &  3.2 &  3.2 &  3.2 &  3.2 &  3.2 \\
{\bfseries B2} &  1.5 &  1.5 &  1.5 &  1.5 &  1.5 &  1.5 &  ---- &
1.0 &  1.0 &  1.0 &  1.0 &  1.0 &  2.4 &  2.4 &  2.4 &  2.4 &  2.4 &
2.4 &  2.4 &  2.4 &  2.4 &  2.4 &  2.4 &  2.4 \\
{\bfseries B2''} &  3.2 &  3.2 &  3.2 &  3.2 &  3.2 &  3.2 &  1.0 &
---- &  1.0 &  1.0 &  1.0 &  1.0 &  1.8 &  1.8 &  1.8 &  1.8 &  1.8 &
1.8 &  1.8 &  1.8 &  1.8 &  1.8 &  1.8 &  1.8 \\
{\bfseries B4}  &  2.4 &  2.4 &  2.4 &  2.4 &  2.4 &  2.4 &  1.0 &
1.0 &  ---- &  1.0 &  1.0 &  1.0 &  2.4 &  2.4 &  2.4 &  2.4 &  2.4 &
2.4 &  1.5 &  1.5 &  1.5 &  1.5 &  1.5 &  1.5 \\
{\bfseries B4'} &  1.8 &  1.8 &  1.8 &  1.8 &  1.8 &  1.8 &  1.0 &
1.0 &  1.0 &  ---- &  1.0 &  1.0 &  1.8 &  1.8 &  1.8 &  1.8 &  1.8 &
1.8 &  3.2 &  3.2 &  3.2 &  3.2 &  3.2 &  3.2 \\
{\bfseries B5}  &  2.4 &  2.4 &  2.4 &  2.4 &  2.4 &  2.4 &  1.0 &
1.0 &  1.0 &  1.0 &  ---- &  1.0 &  1.5 &  1.5 &  1.5 &  1.5 &  1.5 &
1.5 &  2.4 &  2.4 &  2.4 &  2.4 &  2.4 &  2.4 \\
{\bfseries B5'}  &  1.8 &  1.8 &  1.8 &  1.8 &  1.8 &  1.8 &  1.0 &
1.0 &  1.0 &  1.0 &  1.0 &  ---- &  3.2 &  3.2 &  3.2 &  3.2 &  3.2 &
3.2 &  1.8 &  1.8 &  1.8 &  1.8 &  1.8 &  1.8 \\
{\bfseries C1}  &  1.8 &  1.8 &  1.8 &  1.8 &  1.8 &  1.8 &  1.8 &
1.8 &  1.8 &  1.8 &  1.8 &  1.8 &  ---- &  1.0 &  1.0 &  1.0 &  1.0 &
1.0 &  3.2 &  3.2 &  3.2 &  3.2 &  3.2 &  3.2 \\
{\bfseries C1'} &  1.8 &  1.8 &  1.8 &  1.8 &  1.8 &  1.8 &  1.8 &
1.8 &  1.8 &  1.8 &  1.8 &  1.8 &  1.0 &  ---- &  1.0 &  1.0 &  1.0 &
1.0 &  3.2 &  3.2 &  3.2 &  3.2 &  3.2 &  3.2 \\
{\bfseries C3}  &  1.5 &  1.5 &  1.5 &  1.5 &  1.5 &  1.5 &  2.4 &
2.4 &  2.4 &  2.4 &  2.4 &  2.4 &  1.0 &  1.0 &  ---- &  1.0 &  1.0 &
1.0 &  2.4 &  2.4 &  2.4 &  2.4 &  2.4 &  2.4 \\
{\bfseries C3''}  &  3.2 &  3.2 &  3.2 &  3.2 &  3.2 &  3.2 &  1.8 &
1.8 &  1.8 &  1.8 &  1.8 &  1.8 &  1.0 &  1.0 &  1.0 &  ---- &  1.0 &
1.0 &  1.8 &  1.8 &  1.8 &  1.8 &  1.8 &  1.8 \\
{\bfseries C5} &  2.4 &  2.4 &  2.4 &  2.4 &  2.4 &  2.4 &  1.5 &
1.5 &  1.5 &  1.5 &  1.5 &  1.5 &  1.0 &  1.0 &  1.0 &  1.0 &  ---- &
1.0 &  2.4 &  2.4 &  2.4 &  2.4 &  2.4 &  2.4 \\
{\bfseries C5''}  &  1.8 &  1.8 &  1.8 &  1.8 &  1.8 &  1.8 &  3.2 &
3.2 &  3.2 &  3.2 &  3.2 &  3.2 &  1.0 &  1.0 &  1.0 &  1.0 &  1.0 &
---- &  1.8 &  1.8 &  1.8 &  1.8 &  1.8 &  1.8 \\
{\bfseries D1} &  1.8 &  1.8 &  1.8 &  1.8 &  1.8 &  1.8 &  1.8 &
1.8 &  1.8 &  1.8 &  1.8 &  1.8 &  3.2 &  3.2 &  3.2 &  3.2 &  3.2 &
3.2 &  ---- &  1.0 &  1.0 &  1.0 &  1.0 &  1.0 \\
{\bfseries D1''} &  1.8 &  1.8 &  1.8 &  1.8 &  1.8 &  1.8 &  1.8 &
1.8 &  1.8 &  1.8 &  1.8 &  1.8 &  3.2 &  3.2 &  3.2 &  3.2 &  3.2 &
3.2 &  1.0 &  ---- &  1.0 &  1.0 &  1.0 &  1.0 \\
{\bfseries D4}  &  2.4 &  2.4 &  2.4 &  2.4 &  2.4 &  2.4 &  1.5 &
1.5 &  1.5 &  1.5 &  1.5 &  1.5 &  2.4 &  2.4 &  2.4 &  2.4 &  2.4 &
2.4 &  1.0 &  1.0 &  ---- &  1.0 &  1.0 &  1.0 \\
{\bfseries D4''} &  1.8 &  1.8 &  1.8 &  1.8 &  1.8 &  1.8 &  3.2 &
3.2 &  3.2 &  3.2 &  3.2 &  3.2 &  1.8 &  1.8 &  1.8 &  1.8 &  1.8 &
1.8 &  1.0 &  1.0 &  1.0 &  ---- &  1.0 &  1.0 \\
{\bfseries D6} &  1.5 &  1.5 &  1.5 &  1.5 &  1.5 &  1.5 &  2.4 &
2.4 &  2.4 &  2.4 &  2.4 &  2.4 &  2.4 &  2.4 &  2.4 &  2.4 &  2.4 &
2.4 &  1.0 &  1.0 &  1.0 &  1.0 &  ---- &  1.0 \\
{\bfseries D6''} &  3.2 &  3.2 &  3.2 &  3.2 &  3.2 &  3.2 &  1.8 &
1.8 &  1.8 &  1.8 &  1.8 &  1.8 &  1.8 &  1.8 &  1.8 &  1.8 &  1.8 &
1.8 &  1.0 &  1.0 &  1.0 &  1.0 &  1.0 &  ---- \\ \hline

\end{tabular}}}

\caption{Normalized jog-formation energies resulting from
crossings of \bcc dislocations.} \label{tab:matprop3}
\end{table}
\bigskip

A derivation entirely analogous to that leading to
Eq.~(\ref{YieldStressDependence}) yields  
the following expression for the strength of an obstacle 
in the slip system $\alpha$ produced by a forest segment 
in the system $\beta$ 
\begin{equation}
s^{\alpha\beta}= \frac{{s^{\alpha\beta}_0} }{\beta E^{\rm
jog}_{\alpha\beta}} \asinh \left(
\frac{\dot{\gamma}^\alpha}{\dot{\gamma}^\alpha_0} {\rm e}^{\beta
E^{\rm jog}_{\alpha\beta}} \right)
\label{ObstacleStrengthDependence}
\end{equation}
where the strength at zero temperature is given by 
\begin{equation}
s_0^{\alpha\beta} = \frac{E_{\alpha\beta}^{\rm jog}}{b
\bar{l}^{\alpha} L^{\rm junct}}
\end{equation}
and the reference strain rate by
\begin{equation}
\dot{\gamma}^\alpha_0 = 2 \rho^\alpha b \bar{l}^\alpha \nu_{D}
\end{equation}
The lengths $\bar{l}^\alpha$ and $L^{\rm junct}$ describe the geometry
of the junction as illustrated in Fig.~(\ref{fig:junction2}). These
values, which have been estimated to be of the order of few $b$ in the
present case, can also be obtained by atomistic models.  
 
\subsection{Dislocation Evolution: Multiplication and Attrition}
The density of forest obstacles depends directly on the
dislocation densities in all slip systems of the crystal.
Therefore, in order to close the model we require a equation of
evolution for the dislocation densities. Processes resulting in
changes in dislocation density include production by fixed
sources, such as Frank-Read sources, breeding by double cross slip
and pair annihilation (see \cite{kuhlmann:1989} for a review; see
also \cite{johnston:1959, johnston:1960, gillis:1965,
essmann:1973, Lagerlof1993, Dybiec1995}). Although the operation of 
fixed Frank-Read sources is quickly eclipsed by production
due to cross slip at finite temperatures, it is an important 
mechanisms at low temperatures. The double cross-slip, fixed
Frank-Read sources and pair annihilation mechanisms are next 
considered in turn.

\subsubsection{Dislocation Multiplication: Fixed Frank-Reed and Breeding by
  Cross Glide}

The rate of dislocation multiplication in a given slip system $\alpha$
produced by fixed Frank-Reed sources and by breeding by cross glide is
written as
\begin{equation}\label{DCG:RhoDot3}
b \dot{\rho}^\alpha = \lambda_0 \sqrt{\rho^\alpha}\dot{\gamma}^\alpha
\end{equation}
where $\lambda_0$ is a constant associated with the fixed Frank-Read
production; this parameter is rather topological than material dependent.

\subsubsection{Attrition: Pair Annihilation}
The rate of dislocation attrition due to pair annihilation may
finally be estimated as:
\begin{equation}
b\dot{\rho}^\alpha = - \kappa \rho^\alpha \dot{\gamma}^\alpha
\label{DensityAnnihilation}
\end{equation}
where $\kappa$ is the effective annihilation distance. This is the
maximum distance at which two screw segments with opposite direction
and forced to move with a velocity $v = \dot\gamma/b\rho$ will
annihilate. This distance can be estimated by simply equating the
time required for trapping and escaping. Trapping is governed by the
elastic interaction forces (attraction) while escaping by the applied 
strain rate. Then, 
\begin{equation}\label{PA:Kappa}
\frac{1}{\kappa} = \frac{1}{\kappa_c} + \frac{1}{\kappa_0 \left( A + \sqrt{A^2 + 1} \right)}
\end{equation}
where 
\begin{equation}\label{PA:A}
A = {\rm e}^{- \beta E^{\rm jog}} \beta E^{\rm jog} \dot{\gamma}_0^{\rm
jog}/\dot{\gamma}^\alpha 
\end{equation}
is a factor depending on the strain rate and temperature,
\begin{equation}\label{PA:GammaDot0}
\dot{\gamma}_0^{\rm jog} = 2 b \rho l_P \nu_D
\end{equation}
is a reference slip-strain rate and $\kappa_{c}$ is the cut-off value
corresponding the effective screening distance.
It follows that the critical pair-annihilation distance $\kappa$
decreases with increasing strain rate and decreasing temperature.
Thus, at high strain rates the dislocation velocities are high and
the probability of being captured by another dislocation
diminishes accordingly. Additionally, an increase in temperature
increases the dislocation mobility and speeds up the annihilation
process, which results in an attendant increase in annihilation
rates. The rate of annihilation is then modulated by the nucleation energy of
a jog $E^{\rm jog}$ which can be calculated from atomistic simulations.

\section{Atomistic modeling of dislocations properties}  
\label{atomistic}

In the previous section we have identified the following set of {\it 
material parameters} required to estimate the contribution
of each of the controlling unit processes: 
$E^{\text{kink}}$, $L^{\rm kink}$, $U^{\text{edge}}$
$U^{\text{screw}}$, and $E^{\text{jog}}$. In this
section we briefly describe the computation of them using a First Principles-based
Force Field with Molecular Dynamics.

Quantum mechanics (QM) describes the atomic interactions from first principles,
i.e. using no empirical input. Unfortunately QM methods are computationaly
intensive and thus only applicable to small systems (hundreds of atoms)
and short times (picoseconds). The study of most of the
unit processes that govern the plasticity of materials (such as dislocation
mobility, kink energies, etc.) involve many atoms and long simulation
times. Such problems require the use of Force Fields,
where the total energy of the system is given by a potential energy function 
of the atomic positions and does not involve the solution of Shrodinger's equation.
The drawback of using potentials to describe the atomic interactions
is that some accuracy is lost; it is thus of great importance to use accurate
force fields to describe the atomic interactions.

We developed a many body Force Field for Tantalum based on accurate QM 
calculations \cite{strachan2001} which can be used with Molecular Dynamics (MD)
to simulate systems containing millions of atoms. We fitted an Embedded Atom Model type 
Force Field (named qEAM FF) to a variety of ab initio 
calculations, including the zero temperature Equation of State (EOS) for bcc, fcc, 
and A15 phases of Ta in a wide pressure range, elastic constants, vacancy 
formation energy and energetics of a shear transformation in the twinning direction 
\cite{strachan2001}. Ta is a bcc metal and no phase transition to other
crustalline phase is known, but using QM we can calculate the EOS of thermodynamicaly
unstable or metastable phases (such as A15, fcc, hcp, etc.). 
Including data about these
high energy phases, with different coordination numbers, in the Force Field 
training set is important to correctly describe the atomic interactions
near defects such as dislocations, grain boundaries, etc.

We have used the qEAM with MD to study a variety of materials properties such 
as the melting temperature of Ta as a function of pressure \cite{strachan2001}, 
dislocation properties \cite{WangStrachanCaginGoddard2000}, and spall failure \cite{strachan01b}.

\begin{figure}
\centerline{
  \epsfig{file=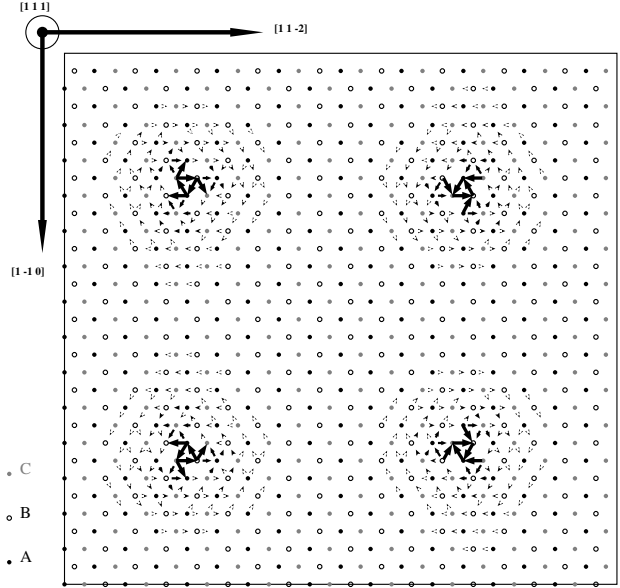,width=0.5\textwidth}
} 
\caption{Differential displacement map of a relaxed quadrupole of screw dislocation in Ta.}
\label{ddmap_screw}
\end{figure}

In subsection \ref{atomistic_corene} we show the calculation of the core energy of edge and screw 
dislocations
in Ta and in Subsection \ref{atomistic_kinks} we calculate the double kink formation energy and 
nucleation length.

\subsection{Core energy of the $1/2 a <111>$ screw and edge dislocations}
\label{atomistic_corene}

    In order to study static properties of the 1/2a$<111>$ screw dislocation in Ta 
such us core structure and energy we use a 
dislocation quadrupole in a simulation cell with periodic boundary conditions. Two of the
dislocations have Burgers vector b=1/2a$<111>$ and the other two have b=-1/2a$<-1-1-1>$. 
Such an arrangement 
of dislocations minimizes the misfit of atoms on the periodic boundary due to the effects of 
periodic images. We build the dislocations using the atomic displacements obtained from elasticity 
theory and then we relax the atomic coordinates using the qEAM FF. In the bcc structure, there are 
two kinds of 
dislocation core configurations (easy core and hard core) that can be transformed to each other by 
reversing the Burgers vector \cite{XuMoriarty1996}. In this work we focus on the lower energy easy cores. 
In Figure \ref{ddmap_screw}
we show the differential displacement map (DDM) of our relaxed quadrupolar system. In the DDM atoms 
are represented by circles and projected on a (111) plane. The arrows represent the relative 
displacement in [111] direction of neighboring atoms due to the dislocation. We can see from 
Figure \ref{ddmap_screw} that the equilibrium dislocation core obtained using qEAM FF has 
three-fold symmetry and 
spreads out in three $<112>$ directions on {110} planes.

\begin{figure}
\centerline{
  \epsfig{file=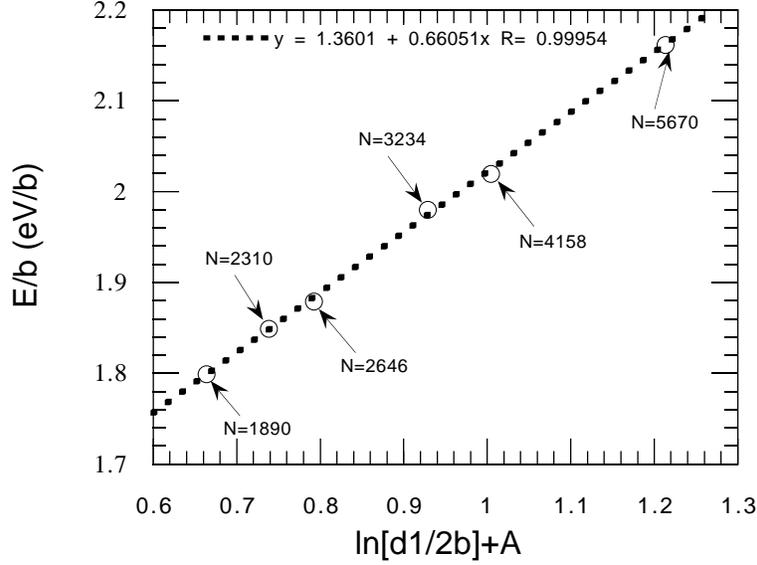,width=0.60\textwidth}
} 
\caption{Total strain energy of the quadrupolar system as a function of  $\ln{d_1/r_c} + A(d_1/d_2)$;
the number of atoms in each simulation is shown. The dashed line is the linear fit to our atomistic
data.}
\label{elastic_ene}
\end{figure}

    Lets define strain energy as the total energy of our system once the perfect crystal energy is 
subtracted. The total strain energy can be divided in two terms: core energy ($E_c$) and elastic 
energy ($E_e$). The latter contains the self-energy of each dislocation and their interactions and can 
be calculated using linear elasticity theory. The core energy is the energy contained close to the 
dislocation line (closer than some distance $r_c$ called core radius) where, due to the large strain, 
elasticity theory is not valid and the details of the interatomic interactions are important. For our 
quadrupole system the total strain energy takes the form \cite{Ismail-beigiArias2000}:
\begin{equation}\label{elast_ene}
E = E_c(r_c) = K b^3 \left[ \ln{\frac{d_1}{r_c}} + A\left(\frac{d_1}{d_2} \right) \right],
\end{equation}
where K depends on the elastic constants, $d_1$ and $d_2$ are the nearest separation of dislocations 
along $<11-2>$ and $<1-10>$  directions and $A(d_1/d_2)$ is a geometric factor which comes from the 
dislocation interactions.

   We studied quadrupolar dislocation cells of different sizes. In Figure (\ref{elastic_ene}) 
we show the minimized 
energy as a function of $\ln{d_1/r_c} + A(d_1/d_2)$ for the different simulation cells; 
in order to compare 
with previous calculations \cite{XuMoriarty1996,Ismail-beigiArias2000} we took $r_c=2b$. We can 
see that the total energies follow a straight 
line as predicted by elasticity theory (Eq. \ref{elast_ene}), showing that the value chosen 
for the core radius 
is large enough to take account for the non elastic region near the dislocation line. 
From a linear fit to our data we determine the core energy $E_c=1.30$ eV/b and 
$K = 3.33\times10^{-2} eV/A3$. 
The value of K can also be computed from the elastic constants giving $3.33\times10^{-2} eV/A3$ 
in excellent agreement
with the one obtained from the fit. Recent ab initio calculations of core energy (using 
periodic cells containing 90 atoms) give 0.86 eV/b, lower 
than the value obtained with qEAM FF and the dislocation cores are compact and symmetric 
\cite{Ismail-beigiArias2000}.

\begin{figure}
\centerline{
  \epsfig{file=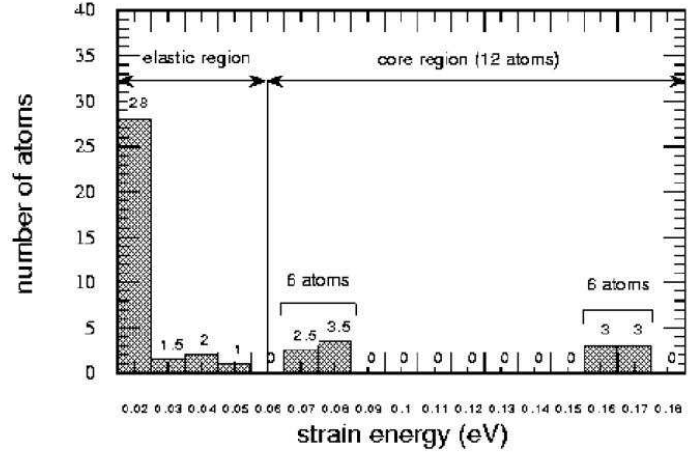,width=0.60\textwidth}
} 
\caption{Histogram of atomistic strain energy distribution for the quadrupolar arrangement of 
screw dislocations. The cell contains 5670 atoms and is 7 Burgers vectors long.}
\label{ene_dist_screw}
\end{figure}

   Using the qEAM we can calculate the strain energy associated with each atom. In Figure 
(\ref{ene_dist_screw}) we show the 
atomic energy distribution (number of atoms per dislocation per Burgers vector as a function of their 
strain energy) for a system containing 5670 atoms in the periodic cell. We can see that there are 6 
atoms with atomic strain energy higher than 0.15 eV and another 6 atoms with energy in the range 
0.06-0.08 eV. They correspond to the 12 atoms per dislocation per Burgers vector closer 
to the dislocation line and their total energy is 1.35eV/b, very similar the core energy 
obtained from Eq. \ref{elast_ene}. The rest of the 
atoms have lower strain energy and can be cosidered as the elastic part of the system. We can then 
define the dislocation core as formed by the 12 atoms per Burgers vector with higher energy.

\begin{figure}
\vglue -0.3\textwidth
\centerline{
  \epsfig{file=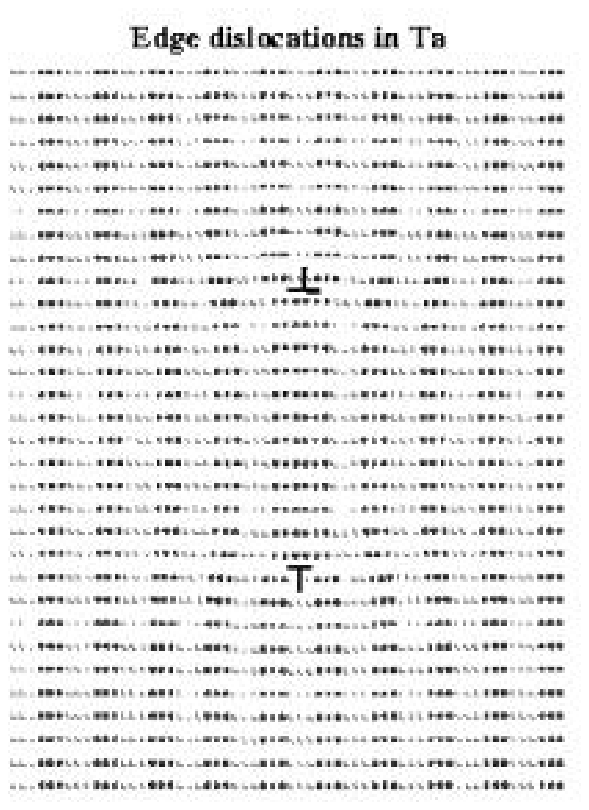,width=1.4\textwidth}
} 
\vglue -0.2\textwidth
\caption{Snapshot of the relaxed edge dipole configuration. The cell
  contains 5652 atoms.}
\label{edge_str}
\end{figure}

We have also calculated the core energy of the edge dislocation with $b=1/2 a <111>$ on a $(110)$ plane.
We build a simulation cell with axis oriented along $<112>$ (x axis), $<110>$ (y axis), and $1/2a<111>$ 
(z axis); this cell contains 6 atoms. We then replicate the cell 3 times along x, 16 times along y, and
20 times along z; the number of atoms in the cell is then N=5760. We then remove 108 atoms to form
a dipole of edge dislocations. Once the system is relaxed (both atoms and cell parameters) we have
a $24.3967$\AA $\times 75.1824$\AA $\times 56.632$~\AA cell. Figure (\ref{edge_str})
shows a snapshot of the atoms projected on a $<112>$ plane. 

\begin{figure}
\vglue -0.2\textwidth
\centerline{
  \epsfig{file=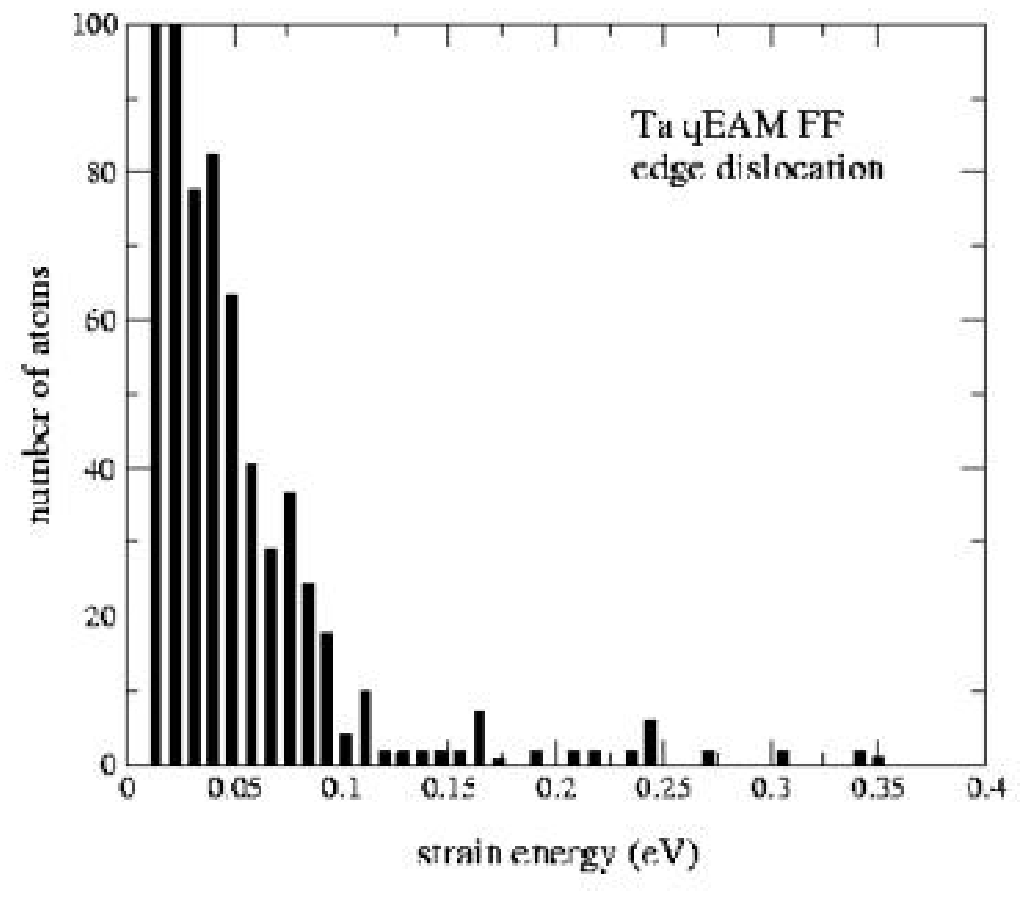,width=1.1\textwidth}
} 
\vglue -0.2\textwidth
\caption{Histogram of atomistic strain energy distribution for the dipole of edge dislocations.
The number of atoms is given per dislocation and per $1/2a<112>$ length.}
\label{edge_ene_dist}
\end{figure}

In Figure \ref{edge_ene_dist} we show the energy distribution for the edge
dislocation (number of atoms per dislocation and per $a<112>$ length as a function of their 
energy). Figure \ref{edge_ene_dist} shows that the core of the edge dislocation contains atoms 
with higher energies
and a broader distribution of energies as compared with the screw case 
[Figure (\ref{ene_dist_screw})].
Taking into accound Figure \ref{edge_ene_dist} we define the core of the edge dislocation as formed 
by those atoms with strain energy higher that 0.1 eV. This definition leads to 36 atoms per 
$a<112>$ or $\sim 4.42$ atoms per \AA~ and to a core energy of $E_{core}^{edge} = 0.827$ eV/\AA
(in the case of the screw we had 12 atoms/b or $\sim4.17$ atoms per \AA). The ratio between the core 
energy of the edge and that of the screw is: $E_{core}^{edge}/E_{core}^{screw} \sim 1.77$. 
It is important to mention that changing the number of atoms considered to belong to the core changes
the core energy, but the difference is minor. Had we taken the 34 atoms per $a<112>$ with higher
energy as the core (leading to $\sim4.18$ atoms /\AA, a density very similar to the one obtained 
in the screw
dislocation) we get a very similar core energy: $E_{core}^{edge} = 0.80$ eV/\AA.

\subsection{Kink pair energy and nucleation lenght}
\label{atomistic_kinks}

As already explained the kink pair mechanism controls the mobility of screw dislocations
in bcc metals and atomistic simulations can provide the details of this mechanism. 

As we can see from Figure (\ref{ddmap_screw}) the core of the screw dislocation spreads
in three $<112>$ directions, this leads to two distint, but energetically equilavent, core
configurations; we will name them positive (P) and negative (N) cores \cite{Wang_kink}.
The shorter (and lower energy) kinks possible involves the displacement of the position of 
the dislocation line in the (111) plane from one equilibrium position to a nearest 
neighbor equilibrium position; the displacement involved is $1/3 a<112>$. There are six
possible $<112>$ directions but only two need to be considered by symmetry, this leads to
two kink directions which we will call left (L) and right (R). The two dislocation 
cores (N and P) and two directions (L and R) lead to 8 different single kinks: 
NRP, NRN, PRP, PRN, NLP, NLN, PLP and PLN.
We have studied all of them in detail \cite{Wang_kink},
here we will concentrate on the single kinks that lead to the lowest energy kink pair.
We calculated the formation energy and length of the various kinks using quadrupolar arrangements
of dislocations as explained in subsection \ref{atomistic_corene}. The simulation cell lengths 
are $40.7$ \AA~ in the 
[11-2] direction, $42.3$ \AA~ in
[1-10] and $431.8$ \AA~ in [111] containing 40500 atoms; the details of these calculations
can be found in \cite{Wang_kink}. We calculate the kink energy as the difference of strain
energy between the
quadrupolar systems containing kinks and that corresponding to perfect straight dislocations. This
energy difference is divided by four to get the energy per kink.
Using the qEAM FF we find that the lowest energy kink pair is formed combining the PLN and NRP kinks.
We define the kink pair nucleation energy as the sum of the formation energy of the
two songle kinks leading to $E^{kink} = 0.725$ eV. The nucleation energy calculated in this way does
not take into account the attractive interaction between the two kinks which lowerers the nucleation
energy. This interation energy is very small for separation larger than $\sim 15$ b 
\cite{XuMoriarty1996,XuMoriarty1998}.

\begin{figure}
\vglue -0.1\textwidth
\centerline{
\hbox{
\hglue -0.1\textwidth
{\epsfig{file=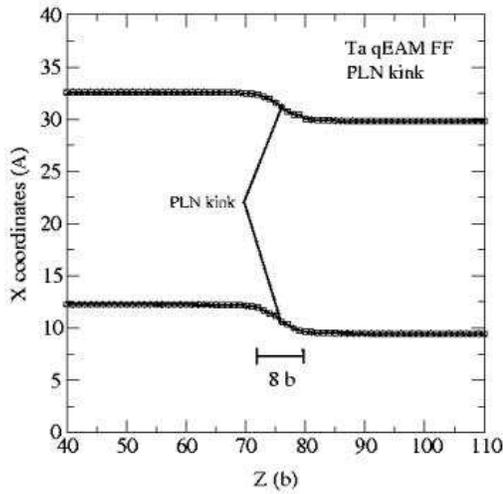,height=0.45\textheight}}\hglue -0.2\textwidth
{\epsfig{file=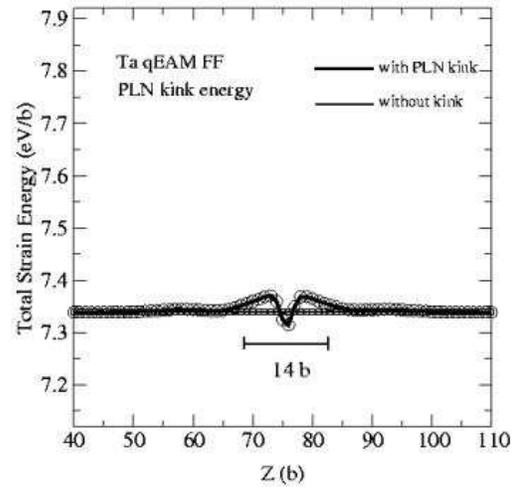,height=0.45\textheight}}
}
}
\vglue -0.1\textwidth
\centerline{
\hbox{
\hglue -0.1\textwidth
{\epsfig{file=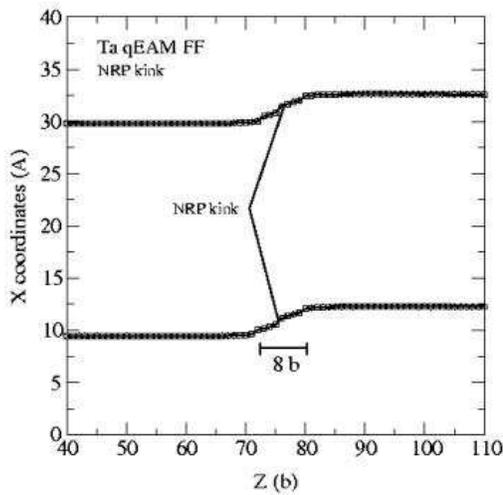,height=0.45\textheight}}\hglue -0.2\textwidth
{\epsfig{file=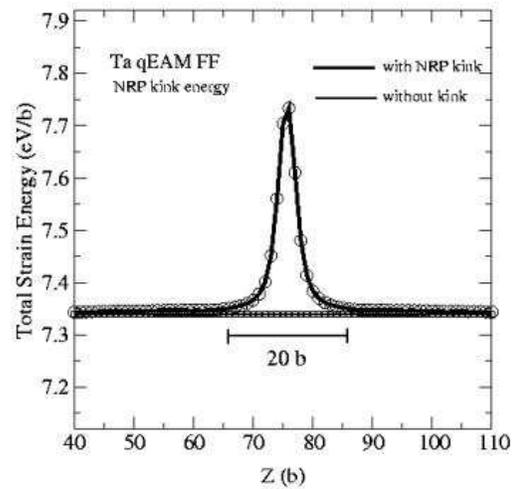,height=0.45\textheight}}
}
}
\vglue -0.1\textwidth
\caption{PLN and NRP kinks in Ta using the qEAM FF. (Top Left) PLN kink: Dislocation position in the
[11-2] direction along the dislocation line; we can see the dislocation moves from one equilibrium
position to the next in a distance of 10 Burgers vectors. (Top Right) PLN kink: total energy in the
quadrupolar system with four PLN kinks along the dislocation line. The system is divided in slices
with thickness equal to b and the energy in each region is
calculated. (Bottom Left) NRP kink: Dislocation 
position in the
[11-2] direction along the dislocation line; we can see the dislocation moves from one equilibrium
position to the next in a distance of 10 Burgers vectors. (Bottom Right) NRP kink: total energy in the
quadrupolar system with four PLN kinks along the dislocation line.  }
\label{kinks}
\end{figure}

As explained above, a critical parameter for the micro-mechanical modeling of plasticity
is, apart from the kink pair energy, its nucleation length $L_{kink}$. We studied both the
energetics and structure of the variuos kinks
along the dislocation line.  Figure \ref{kinks} shows
the extent of the kinks both from structural and energetic points of view. 
We show the position of the dislocation
in the direction of the kink along the dislocation line for a PLN kink [Figure \ref{kinks}(a)] and
NRP kink [Figure \ref{kinks}(c)]. We also show the total energy of the quadrupolar system along the
dislocation line for the PLN [Figure \ref{kinks}(b)] and NRP [Figure \ref{kinks}(d)] kinks.
This is calculated by dividing the system in the [111] direction in regions of width equal to
the Burgers vector and calculating the total energy in each slice. The structural length
of the PLN kinks is $L^{PLN}_{str} = 8$ b [Figure \ref{kinks} (a)]; while its ``energetic extent'' 
is  $L^{PLN}_{ene} = 14$ b [Figure \ref{kinks} (b)].
For NRP kinks we obtain: $L^{NRP}_{str} = 8$ b [Figure \ref{kinks} (c)] and 
$L^{NRP}_{ene} = 20$ b [Figure \ref{kinks} (d)]. 

\begin{figure}
\vglue -0.1\textwidth
\centerline{
  \epsfig{file=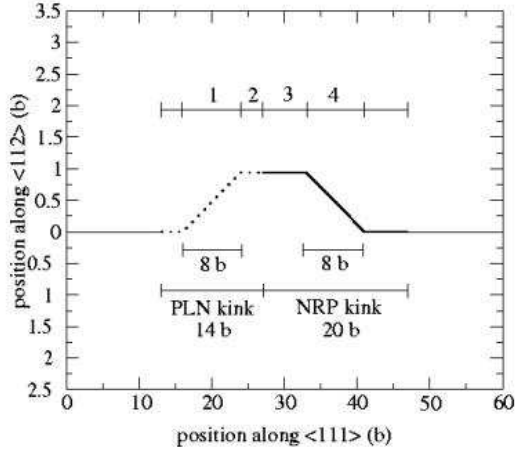,width=0.8\textwidth}
} 
\vglue -0.1\textwidth\caption{Schematic diagram of a kink pair formed by a NRP and PLN single kinks.
The four terms entering in the work expresion (Equation \ref{work_kink}) are
shown in the Figure.}
\label{kink_scheme}
\end{figure}

Going back to the definitions
of the paramenters entering the equation that governs the dislocation mobility [Equations
(\ref{YieldStressDependence}) and (\ref{tauzero})]; the effective Peierls stress 
($\tau_0$), Equation (\ref{tauzero}), is defined as the applied stress for which the nucleation free
energy for a kink pair ($\Delta G$) is zero. $\Delta G$ is given by:
\begin{equation}
\label{deltag_kink}
\Delta G = E^{kink} \pm \tau l_P b L^{kink},
\end{equation}
where $L^{kink}$ is the effective kink pair nucleation length and $l_P$ is distance advanced by the
dislocations; in the kinks studied here $l_P = | 1/3 a<112> |$.
The second term in the right hand side of Equation \ref{deltag_kink} is the work done
by the external stress when the kink is nucleated. 
Figure \ref{kink_scheme} shows a schematic diagram of a PLN-NRP kink pair. We
can see that the work done by the external stress to nucleate the kink pair can be divided
in four terms:
\begin{equation}
\label{work_kink}
\tau b l_P L^{kink} =  \tau b l_P \left( \frac{L^{PLN}_{str}}{2} + \frac{L^{PLN}_{ene} - L^{PLN}_{str}}{2} +
\frac{L^{NRP}_{ene} - L^{NRP}_{str}}{2} + \frac{L^{NRP}_{str}}{2} \right)
\end{equation}
where $L^{kink}$ is the effective kink pair length. In Figure \ref{kink_scheme} we show the 
four terms in the right hand side of Equation (\ref{work_kink}). Note that Equation \ref{work_kink} 
assumes that the kinks are straight lines connecting the two equilibrium positions of the dislocation.
In this way we obtain the effective kink pair nucleation length $L^{kink} = 17$ b.

The remaining materials parameter is the nucleation energy of a jog $E^{\text{jog}}$.
In this work we take $E^{\text{jog}}$ as the PLN-NRP kink pair nucleation energy.

\section{Experiment, Validation and Prediction}
\label{ComparisonExperiment}

To test the predictive capabilities of the multiscale approach we
first select a set of material parameters to best fit the experimental
results, then we compare these parameters against the atomistically
computed ones and finally we {\it predict} the macroscopic response 
using the atomistics parameters. As we shall see, the agreement
between the {\it fitted} and {\it computed by atomistics} material
parameters is remarkable, and the macroscopic predicted response 
retains most of the experimental features. These facts provide
confidence in the multiscale modeling approach, indicating that even 
in the case that experimental data would not have been available, still 
the macroscopic behavior could have been predicted based only on atomistic 
calculations.

The experiment data correspond to uniaxial tests on Ta single crystals
of Mitchell and Spitzig \cite{mitchell:1965}. In these
tests, 99.97\%-pure Ta specimens were loaded in tension along the
$[213]$ crystallographic axis, at various combinations of temperature
and strain rate. In particular we considered temperatures ranging from
$296 \; K$ to $573 \; K$, and strain rates ranging from 
$10^{-1} \; s^{-1}$ to $10^{-5} \; s^{-1}$. The
numerical procedure employed for the integration of the
constitutive equations has been described elsewhere
\cite{OrtizStainier1999}. The constitutive update is fully
implicit, with the active systems determined iteratively so as to
minimize an incremental work function. All stress-strain curves
are reported in terms of nominal stress and engineering strain.

\begin{figure}
  \centerline{
    \epsfig{file=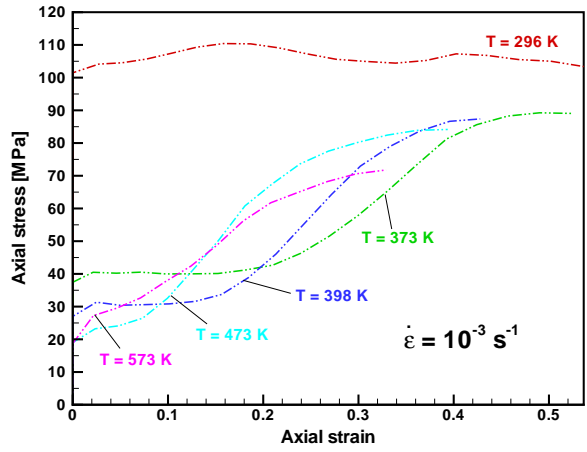,width=0.55\textwidth}
  }
  \centerline{(a) Experimental data of Mitchell \& Spitzig
                  \cite{mitchell:1965}} 
  \centerline{
    \epsfig{file=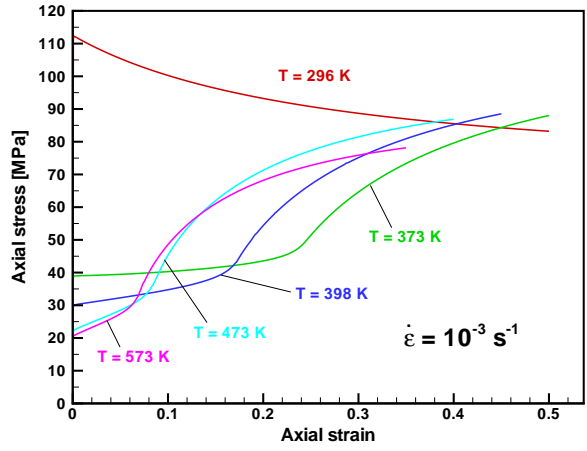,width=0.55\textwidth}
  }
  \centerline{(b) Predictions of the model with fitted parameters}
  \centerline{
    \epsfig{file=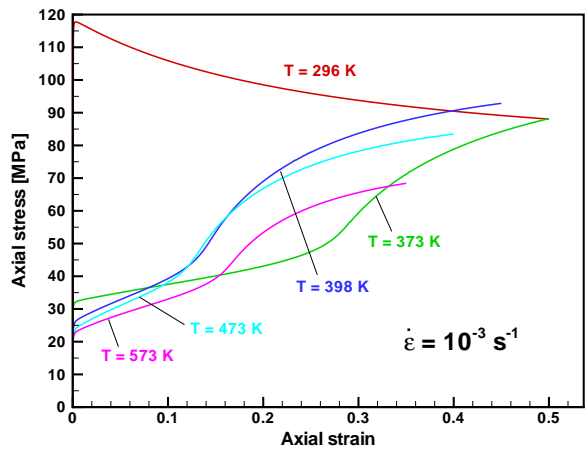,width=0.55\textwidth}
  }
  \centerline{(c) Predictions of the model with atomistic parameters}
  \caption{Temperature dependence of stress-strain curves for
  $[213]$ Ta single crystal
  ($\dot{\epsilon}=10^{-3}$ s${}^{-1}$).}
  \label{fig:stress-strain1}
\end{figure}
\begin{figure}
  \centerline{
    \epsfig{file=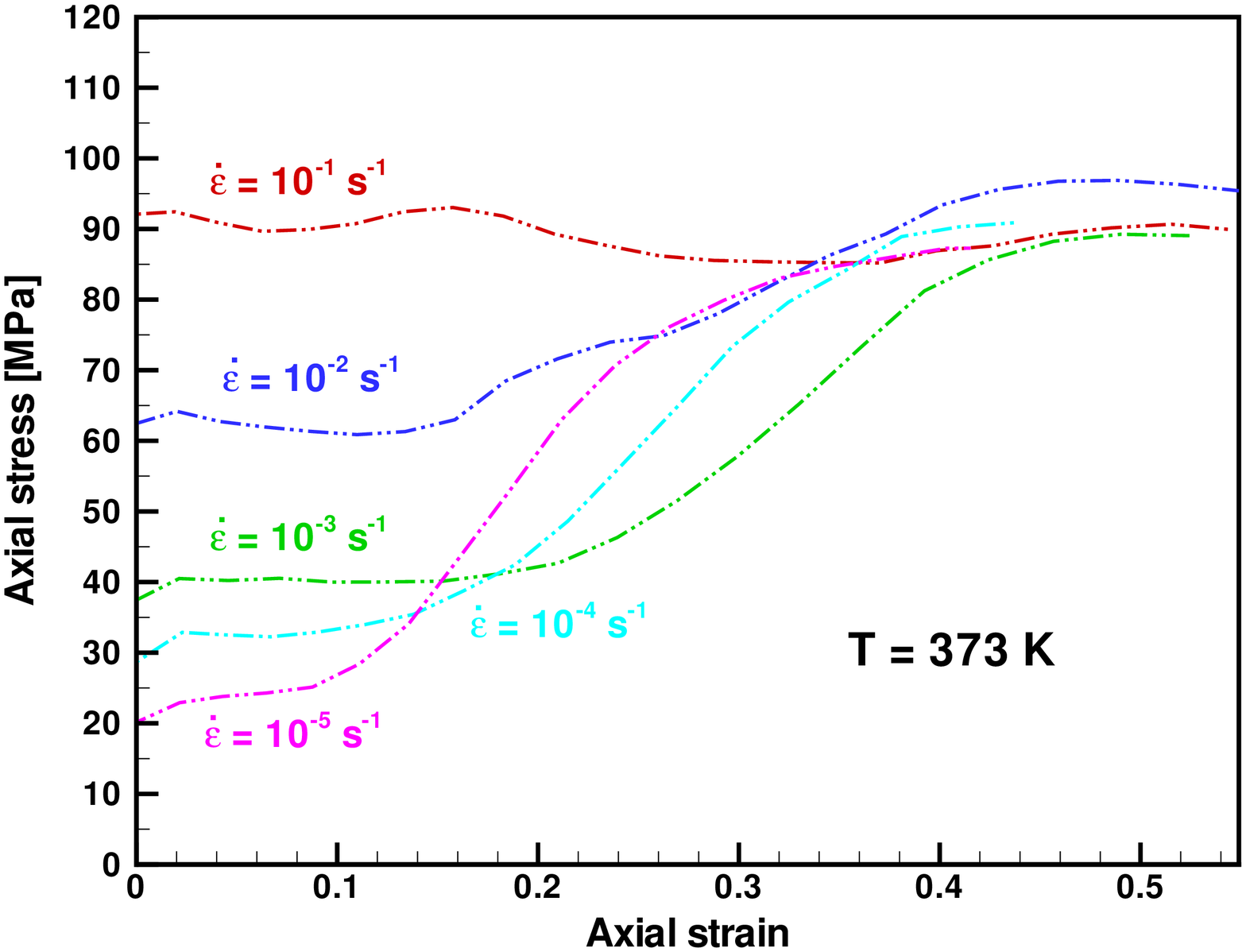,width=0.55\textwidth}
  }
  \centerline{(a) Experimental data of Mitchell \& Spitzig
                  \cite{mitchell:1965}} 
  \centerline{
    \epsfig{file=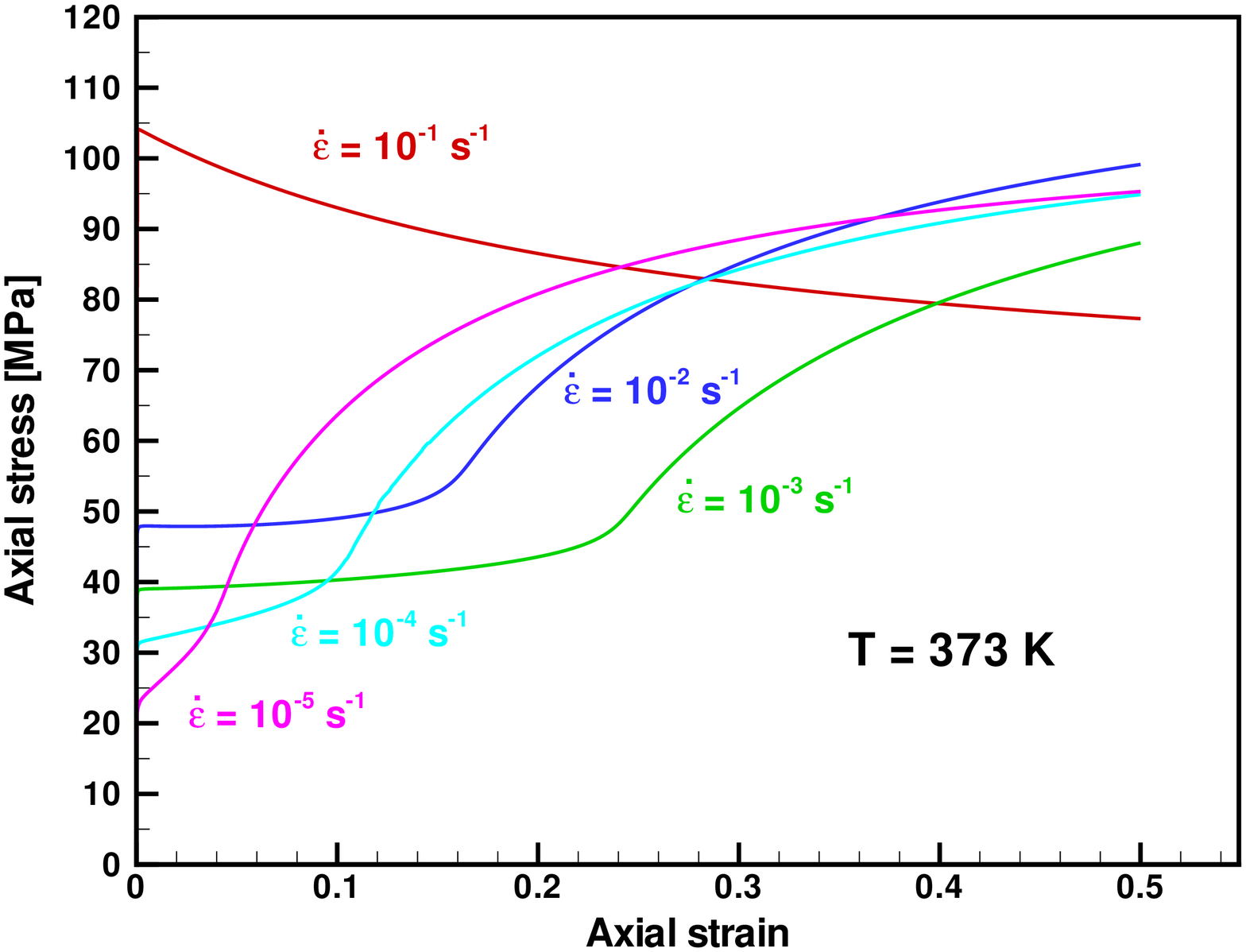,width=0.55\textwidth}
  }
  \centerline{(b) Predictions of the model with fitted parameters}
  \centerline{
    \epsfig{file=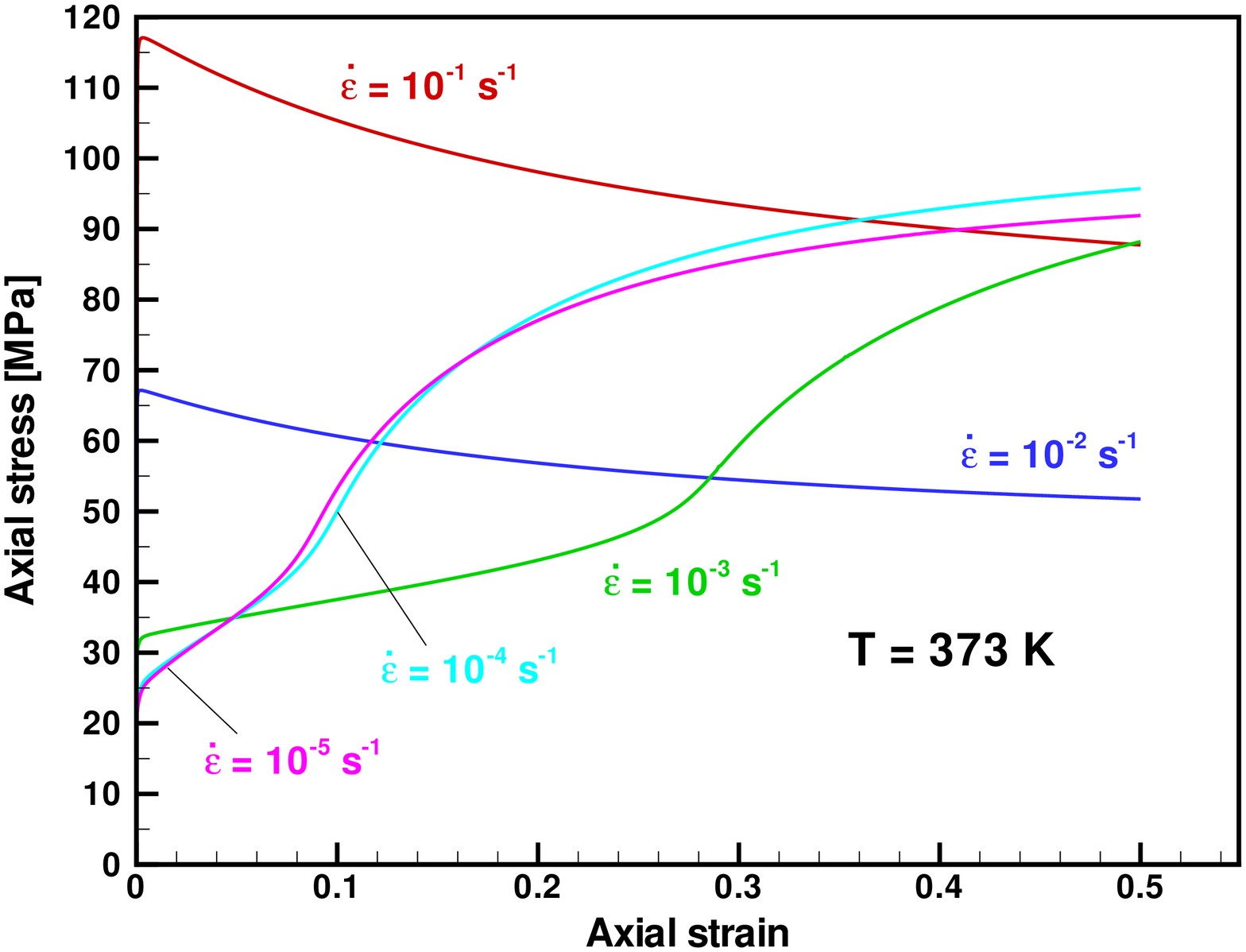,width=0.55\textwidth}
  }
  \centerline{(c) Predictions of the model with atomistic parameters}
  \caption{Strain-rate dependence of stress-strain curves for
  $[213]$ Ta single crystal ($T$=373 K).}
  \label{fig:stress-strain2}
\end{figure}

Two different sets of material properties were used for the numerical
simulations. The first set was obtained by fitting the simulation
results to the experimental results.
Table \ref{tab:properties} identifies the subset of parameters which
are also amenable to direct calculation by atomistic based methods. The
table lists the parameter values obtained by these methods, as
described in the previous sections, in parallel with the values obtained
by the fitting approach. Thus, in the second set of properties which
was used for numerical simulations,
atomistic-based values replace fit-based values, when available. This
is the case for the edge and screw dislocation self-energies, as well as the
kink-pair formation energy and length. Clearly, those two sets don't
differ by much, which strongly support the validity of the advertised
multiscale paradigm. For a complete list of parameters for the model,
the reader should refer to \cite{StainierCuitinoOrtiz2001}.

\begin{table}
  \caption{Material parameters for Tantalum}
  \label{tab:properties}
  \smallskip
  \renewcommand{\arraystretch}{1.2}
  \centering
  \begin{tabular}{cccc}
   \hline
    Parameter           & Fitted set             & Atomistic set    & Units \\
   \hline
    $E^{\text{kink}}$   & $0.70$                 & $0.725$          & [eV] \\
    $L^{\rm kink}/b$    & $13$                   & $17$             & \\
    $U^{\text{edge}}/\mu b^{2}$ $({}^{\footnotesize *})$ & $0.2 $
                                                 & $0.216$          & \\
    $U^{\text{edge}}/U^{\text{screw}}$ & $1.77^{\footnotesize **}$ & $1.77$           & \\
   \hline
    $\bar{l}/b$         & $5$                    & $5$              & \\
    $L^{\rm junct}/b$   & $20$                   & $20$             & \\
   \hline
    $E^{\text{cross}}$  & $0.67$                 & $.725$           & [eV] \\
   \hline
    $\lambda_{\rm FR}$  & $2.3$                  & $4.5^{\footnotesize ***}$
                                                                    & \\
    $\kappa_c/b$        & $1250$                 & $500^{\footnotesize ***}$
                                                                    & \\
   \hline
    \multicolumn{3}{l}{\rule{0pt}{2.5ex} ${}^{\footnotesize *}$
         $\mu=\frac{3}{5} C_{44} + \frac{1}{5} (C_{11}-C_{12})$.}
         \\
    \multicolumn{3}{l}{\rule{0pt}{2.5ex} ${}^{\footnotesize **}$
         Taken from the atomistic simulations.}
         \\
    \multicolumn{3}{l}{\rule{0pt}{2.5ex} ${}^{\footnotesize ***}$
         Not computed by atomistics.}
         \\
  \end{tabular}
\end{table}


Figs.~\ref{fig:stress-strain1} and \ref{fig:stress-strain2} show the
predicted and measured stress-strain curves for a $[213]$ Ta crystal
over a range of temperatures and strain rates. One can compare, from
top to bottom: the experimental results, the results obtained after
fitting the parameters, and the results obtained with atomistic-based
parameters. It is evident from these figures that the model, with both
sets of parameters, captures salient features of the behavior of Ta
crystals such as: the dependence of the initial yield point on
temperature and strain rate; the presence of a marked stage I of easy
glide, specially at low temperature and high strain rates; the sharp
onset of stage II hardening and its tendency to shift towards lower
strains, and eventually disappear, as the temperature increases or the
strain rate increases; the parabolic stage II hardening at low strain
rates or high temperatures; the stage II softening at high strain
rates or low temperatures; the trend towards saturation at high
strains; and the temperature and strain-rate dependence of the
saturation stress. Thus, the predictive approach based on atomistic
methods clearly shows its capacity to produce results matching the
experimental evidence.

The theory reveals useful insights into the mechanisms underlying
these behaviors. For instance, since during state I the crystal
deforms in single slip and the secondary dislocation densities are
low, the Peierls resistance dominates and the temperature and
strain-rate dependency of yield owes mainly to the thermally
activated formation of kinks and crossing of forest dislocations.
It is interesting to note that during this stage the effect of
increasing (decreasing) temperature is similar to the effect of
decreasing (increasing) strain rate, as noted by Tang {\it et al.}
\cite{TangDevincreKubin1999}. The onset of stage II is due to the
activation of secondary systems. The rate at which these secondary
systems harden during stage I depends on the rate of dislocation
multiplication in the primary system. This rate is in turn
sensitive to the saturation strain $\gamma^{\rm sat}$, which
increases with strain rate and decreases with temperature. As a
result, the length of the stage I of hardening is predicted to
increase with strain rate and decrease with temperature, as
observed experimentally. Finally, the saturation stress is mainly
governed by the forest hardening mechanism and, in particular, by
the strength of the forest obstacles. This process is less
thermally activated than the Peierls stress, since the
corresponding energy barriers are comparatively higher.
Consequently, the stress-strain curves tend to converge in this
regime, in keeping with observation.

The apparent softening observed in simulation results at the lowest
temperature ($296\;K$) and the highest strain rate ($10^{-1}\;s^{-1}$)
is actually an effect of the boundary conditions, allowing some level
of rotation of the specimen. Since in those cases, the material
hardening is relatively low (stage I only), this geometrical softening
dominates in the apparent macroscopic behavior. In the other cases,
the activation of several systems at high strains results in a more
isotropic deformation, in turn leading to limited rotations. In order
to take the exact experimental boundary conditions into account, a
finite element model of the whole specimen should be used, allowing
for a non-homogeneous deformation field.

\section*{Acknowledgments}
The support of the DOE through Caltech's ASCI Center for the
Simulation of the Dynamic Response of Materials is gratefully
acknowledged. LS also wishes to acknowledge support from the
Belgian National Fund for Scientific Research (FNRS).

\bibliographystyle{unsrt}

\end{document}